\documentclass[superscriptaddress,nofootinbib,showpacs,preprint,preprintnumbers,amsmath,amssymb]{revtex4-1}
\usepackage{amsmath}
\usepackage{epsfig}
\usepackage{graphicx,subfigure}
\usepackage{color}
\usepackage{latexsym}
\usepackage{amssymb}
\usepackage{enumerate}
\usepackage{hyperref}

\newcommand{\sgh}{\hat{\sigma}}

\newcommand{\lsim}{\lesssim}

\newcommand{\beq}{\begin{equation}}
\newcommand{\eeq}{\end{equation}}
\newcommand{\bea}{\begin{eqnarray}}
\newcommand{\eea}{\end{eqnarray}}
\newcommand{\barr}{\begin{array}}
\newcommand{\earr}{\end{array}}
\newcommand{\bc}{\begin{center}}
\newcommand{\ec}{\end{center}}
\newcommand{\bit}{\begin{itemize}}
\newcommand{\eit}{\end{itemize}}
\newcommand{\ben}{\begin{enumerate}}
\newcommand{\een}{\end{enumerate}}

\newcommand{\sh}{\hat{s}}

\newcommand{\no}{\nonumber}
\newcommand{\nn}{\nonumber}

\newcommand{\br}{{\rm Br}}
\newcommand{\ie}{\textit{i.e.}, }

\newcommand{\fb}{{\,{\rm fb}}}

\newcommand{\gev}{{\;{\rm GeV}}}
\newcommand{\tev}{{\;{\rm TeV}}}

\newcommand{\sg}{\sigma}

\newcommand{\Gm}{\Gamma}

\newcommand{\tb}{t_\beta}

\newcommand{\cba}{c_{\beta-\alpha}}
\newcommand{\sba}{s_{\beta-\alpha}}
\newcommand{\yh}{\hat{y}}

\newcommand{\cphi}{{\,\cos\phi}}
\newcommand{\sphi}{{\,\sin\phi}}

\newcommand{\shat}{{\hat s}}

\newcommand{\dtc}{{\Delta C}}

\newcommand{\rr}{{\gamma \gamma}}
\newcommand{\ttop}{{t\bar{t}}}
\newcommand{\ttbar}{t{\bar t}}

\newcommand{\ttau}{{\tau \tau}}

\newcommand{\Eq}[1]{Eq.~(\ref{#1})}

\newcommand{\calA}{{\cal A}}

\def\nn{\nonumber}

\def\GeV{\,{\rm GeV}}
\def\TeV{\,{\rm TeV}}
\def\fb{\,{\rm fb}}
\def\invfb{\,{\rm fb}^{-1}}

\def\rr{\gamma\gamma}
\def\ttbar{t\bar t}
\def\tb{t_\beta}
\def\cosbma{c_{\beta-\alpha}}
\def\br{{\rm Br}}

\begin{document}
\baselineskip 3.5ex

\title{
Interference effect on heavy Higgs resonance signal \\ in $\rr$ and $ZZ$ channels
}

\author{Jeonghyeon Song}
\email{jeonghyeon.song@gmail.com}
\affiliation{School of Physics, KonKuk University, Seoul 143-701, Korea}
\author{Yeo Woong Yoon}
\email{ywyoon@kias.re.kr}
\affiliation{School of Physics, KonKuk University, Seoul 143-701, Korea}
\affiliation{School of Physics, Korea Institute for Advanced Study, Seoul 130-722, Korea}
\author{Sunghoon Jung}
\email{shjung@slac.stanford.edu}
\affiliation{School of Physics, Korea Institute for Advanced Study, Seoul 130-722, Korea}
\affiliation{SLAC National Accelerator Laboratory, Menlo Park, CA 94025, USA \vspace{0.5cm}}

\begin{abstract}
\vspace{0.3cm}
\baselineskip 4.0ex
The resonance-continuum interference is usually neglected
when the width of a resonance is small compared to the resonance mass.
We re-examine this standard by
studying the interference effects
in high-resolution decay channels, $\rr$ and $ZZ$,
of the heavy Higgs boson $H^0$
in nearly aligned two-Higgs-doublet models.
For the $H^0$ with a sub-percent width-to-mass ratio, we find that, in the parameter space where the LHC 14 TeV $ZZ$ resonance search can be sensitive, the interference effects can modify the $ZZ$ signal rate by ${\cal O}(10)\%$ and the exclusion reach by ${\cal O}(10)$ GeV.
In other parameter space where the $ZZ$ or $\rr$ signal rate is smaller, the LHC 14 TeV reach is absent, but a resonance shape can be much more dramatically changed. In particular, the $\rr$ signal rate can change by ${\cal O}(100)\%$. Relevant to such parameter space, we suggest variables that can characterize a general resonance shape.
We also illustrate the relevance of the width on the interference by adding non-standard decay modes of the heavy Higgs boson.

\end{abstract}

\preprint{KIAS-P15055, SLAC-PUB-16434}

\maketitle

\section{Introduction}

Needless to say, the 125 GeV Standard Model (SM) Higgs boson discovery at the LHC Run I~\cite{Higgs:discovery:2012} is a big step toward the understanding of the electroweak symmetry breaking. But the observed mass of 125 GeV requires a satisfactory explanation for the huge hierarchy between the weak scale and the Planck scale. Most candidate explanations, such as supersymmetry and composite Higgs models, predict a set of new particles at around the electroweak scale. The absence of any such discovery at the LHC Run I motivates us not only to re-ponder naturalness criteria but also to re-visit collider search strategies.

The 13 TeV LHC Run II, which started taking data a few months ago, may indeed need a  careful study of resonance searches. Unlike usually assumed, a particle somewhat heavier than the electroweak scale may not show up as a  Breit-Wigner (BW) resonance peak at the LHC experiments. The resonance-continuum interference can induce observable impacts on the production rate and the invariant mass distribution (resonance shape). It is generally because (a) a heavier particle can be broader (more decay channels with less phase-space suppression and possible Goldstone enhancements), and (b) the production and decay amplitudes can involve complex phases that arise from SM particles running in loops below the threshold. Various studies have shown that the interference for such cases is not usually negligible~
\cite{Jung:2015gta,Gaemers:1984sj,Dicus:1994bm,Morris:1993bx,Niezurawski:2002jx,
Asakawa:1999gz,Dixon:2008xc,Dixon:2003yb,Campbell:2011cu,Dixon:2013haa,
Kauer:2015hia,Glover:1988rg,Bernreuther:1998qv,Basdevant:1992nb,Dicus:1987fk,
Bian:2015hda,Kauer:2012hd,Kauer:2013qba,Englert:2015zra,Kauer:2015dma,
Martin:2012xc,Ellis:2004hw,Farina:2015dua,Bonvini:2013jha}.

Most resonance searches at collider experiments model a resonance as a BW peak and estimate the signal rate by the narrow width approximation (NWA). This is justified if the width-to-mass ratio $\Gamma/M$ is small enough (see e.g. Ref.~\cite{Beneke:2003xh}) and the resonance width is smaller than the experimental resolution. Thus, LHC searches assume 1\% $\Gm/M$ in the $\rr$ channel~\cite{ATLAS:diphoton:resonance:2014,CMS:2014onr} and 0.5\% in the $ZZ$ channel~\cite{Aad:2015kna,Chatrchyan:2013mxa}, which imply that the width of a few hundred GeV resonance is similar or smaller than the experimental bin size. But for even a slightly broader resonance, perhaps with some complex phases in its production and decay amplitudes, such approximation may not be guaranteed. In this paper, we re-examine such approximation using the two-Higgs-doublet models (2HDM).

A notable example that reveals the dramatic interference effect
is the decay of heavy Higgs bosons $H^0$ and $A^0$
into the $\ttbar$ at hadron colliders~\cite{Dicus:1994bm}. Most strikingly, it was shown that a pure resonance dip is produced in a large part of parameter space~\cite{Jung:2015gta}. In other parameter space, a general mixture of the real- and the imaginary-part interferences produces a mixture of a peak and a dip in the $m_{\ttop}$ distribution~\cite{Jung:2015gta,Craig:2015jba,Bernreuther:1997gs}. Unfortunately, it is difficult to resolve such rich structure of the $\ttop$ resonance shape~\cite{Craig:2015jba} because
the experimental $m_\ttop$ resolution $\sim 100$ GeV~\cite{ATLAS:2015aka,Chatrchyan:2013lca} is bigger than the typical width of the heavy Higgs bosons in the aligned 2HDM. Although a pure dip can perhaps be well searched using the available techniques optimized for a BW peak~\cite{Jung:2015gta}, it is produced only in some part of the parameter space.

The interference also exists in the high-resolution decay channels, $\rr$ and $ZZ$.
The interferences of the SM-like heavy Higgs boson at hadron colliders, $gg\to H \to ZZ, \, \rr$, have been calculated in last decades~\cite{Niezurawski:2002jx,Martin:2012xc,Dicus:1987fk,Dixon:2003yb,Kauer:2012hd,Kauer:2013qba,Kauer:2015hia,Englert:2015zra,Kauer:2015dma}; but they are found to be insignificant producing mostly a BW peak. The main difference between the $\ttop$ and $ZZ, \rr$ channels is the relative size of the continuum and the resonance processes~\cite{Jung:2015gta}. The $\ttop$ experiences a significant interference because the tree-level continuum, $B$, and the one-loop resonance, $S$, can produce a loop-factor enhanced interference $\sqrt{SB}/S \sim \sqrt{B/S}$ relative to the resonance-squared.
Large interference effect also expected in $gg\to\rr$ since the continuum is one-loop process while the resonance is via two-loop.
On the other hand, both the $gg\to ZZ$ continuum and the $gg \to H \to ZZ$ resonance processes are at the same one-loop order, so that relatively small interference is expected.
Meanwhile, the off-shell interference of the SM Higgs, which is beyond the scope of this paper, with the continuum $ZZ$ at an invariant mass much bigger than 125 GeV was found to be  sensitive to the Higgs width~\cite{Kauer:2012hd,Caola:2013yja,Khachatryan:2014iha}.

Nonetheless, the interference in the $ZZ$ and $\rr$ can be more exciting for the 2HDM heavy Higgs bosons. The expectation is again based on a general estimation of the relative interference~\cite{Jung:2015gta}. In the nearly-aligned 2HDM, as is preferred by SM Higgs precision measurements ($|c_{\beta-\alpha}| \lesssim 0.1 -0.4$ depending on models), the resonance process is suppressed by the small  $c_{\beta-\alpha}$ and complex phases can be different in the $\rr$ channel as the $W$ boson loop is suppressed. As a result, the interference can be relatively enhanced and the resonance shape can be non-trivially modified. Thus, we study the interference in the $ZZ$ and $\rr$ channels in this nearly-aligned 2HDM. This setup is not only motivated by Higgs precision measurements, but can illustrate the resonance-continuum interference of a relatively narrow resonance.

The paper is organized as follows.
In Sec.~\ref{sec:review}, we review the 2HDM and our formalism of the interference effect on the invariant mass distribution. In addition, we suggest a few variables that can characterize a resonance shape containing a peak and a dip.
We then present the results for the $\rr$ channel in Sec.~\ref{sec:rr} and the $ZZ$ channel in Sec.~\ref{sec:ZZ}.
We also consider the case of a somewhat broader heavy Higgs boson in Sec.~\ref{sec:extra}, in which we add non-standard decay modes.
Section \ref{sec:conclusions} contains our conclusions.

\section{Brief review of the 2HDM and the interference effects}
\label{sec:review}
\subsection{$H^0$ in the 2HDM}
We consider a 2HDM~\cite{2HDM} with \textit{CP} invariance
and softly broken $Z_2$ symmetry, which introduces
two complex Higgs doublet scalar fields, $\Phi_1$ and $\Phi_2$.
The electroweak symmetry breaking is generated by
non-zero vacuum expectation value $v$ of a linear combination $H_1 = \cos\beta \Phi_1 + \sin\beta \Phi_2$.
Its orthogonal combination $H_2 = -\sin\beta \Phi_1 + \cos\beta \Phi_2$
acquires zero vacuum expectation value.
In what follows,  we take $s_x=\sin x$, $c_x = \cos x$, and $t_x = \tan x$
for notational simplicity.
There are five physical Higgs boson degrees of freedom,
the light CP-even scalar $h^0$,
the heavy CP-even scalar $H^0$, the CP-odd pseudoscalar $A^0$,
and two charged Higgs bosons $H^\pm$.
The SM Higgs field
is a mixture of $h^0$ and $H^0$ as
\bea
H^{\rm SM} = \sba h^0 + \cba H^0,
\eea
where $\alpha$ is the mixing angle between $h^0$ and $H^0$.
Note that if $\sba=1$, $h^0$ has the same couplings as the SM Higgs boson,
which is preferred by the SM Higgs precision measurement with LHC8 data~\cite{2hdm:Higgs:fit}.
This is called the alignment limit~\cite{alignment}.

We consider the case where the observed 125 GeV state $h_{125}$
is the lighter \textit{CP}-even state $h^0$
although another interesting possibility of $h_{125}=H^0$ is still compatible with
the current LHC Higgs data~\cite{Wang:2014lta,Kanemura:2014dea,Bernon:2014nxa,Coleppa:2013dya,deVisscher:2009zb,Ferreira:2014dya,Chang:2015goa}.
In addition, we assume $\sba>0$\footnote{
\baselineskip 3.5ex
Note that
the wrong sign case in the 2HDM is shown to be
still allowed by the current LHC Higgs signal strength measurements,
though less probable~\cite{Ferreira:2014dya}.}.
Focused on $\rr$ and $ZZ$ decay modes,
we study the gluon fusion production of $H^0$
in the two-dimensional parameter space of $M_{H}$ and $\tb$ with the given $\cosbma$.
Another model parameter,
the soft $Z_2$ symmetry breaking term $m_{12}^2$,
is tuned to suppress $H^0$-$h^0$-$h^0$ triple coupling.

The $H$-$V$-$V$ ($V=W^\pm,Z^0$) coupling normalized by the SM value is $\cba$.
In order to have $H \to Z Z$,
therefore, we need some deviation from the alignment limit.
The Yukawa couplings are different according to types of 2HDM.
In this study, we consider Type I and Type II
where the normalized Yukawa couplings by the SM values, $\yh^{H}_{t,b,\tau}$,
in terms of $\cba$ and $\sba$ are
\bea
\label{eq:Yukawa:couplings}
\renewcommand*{\arraystretch}{1.2}
\begin{array}{l|cc}
                & ~~~~~\cba - \dfrac{\sba}{\tb}~~~ & ~~~\cba+\tb\sba~~~~~\\
                \hline
\hbox{Type I } & \yh^H_{t} , \yh^H_{b},\yh^H_{\tau} & \\
\hbox{Type II } & \yh^H_{t} & \yh^H_{b},\yh^H_{\tau} \\
\end{array}
\eea
Note that both Type I and Type II have the same top quark Yukawa coupling.

\begin{figure}[t]
\centering
\includegraphics[width=0.52\textwidth]{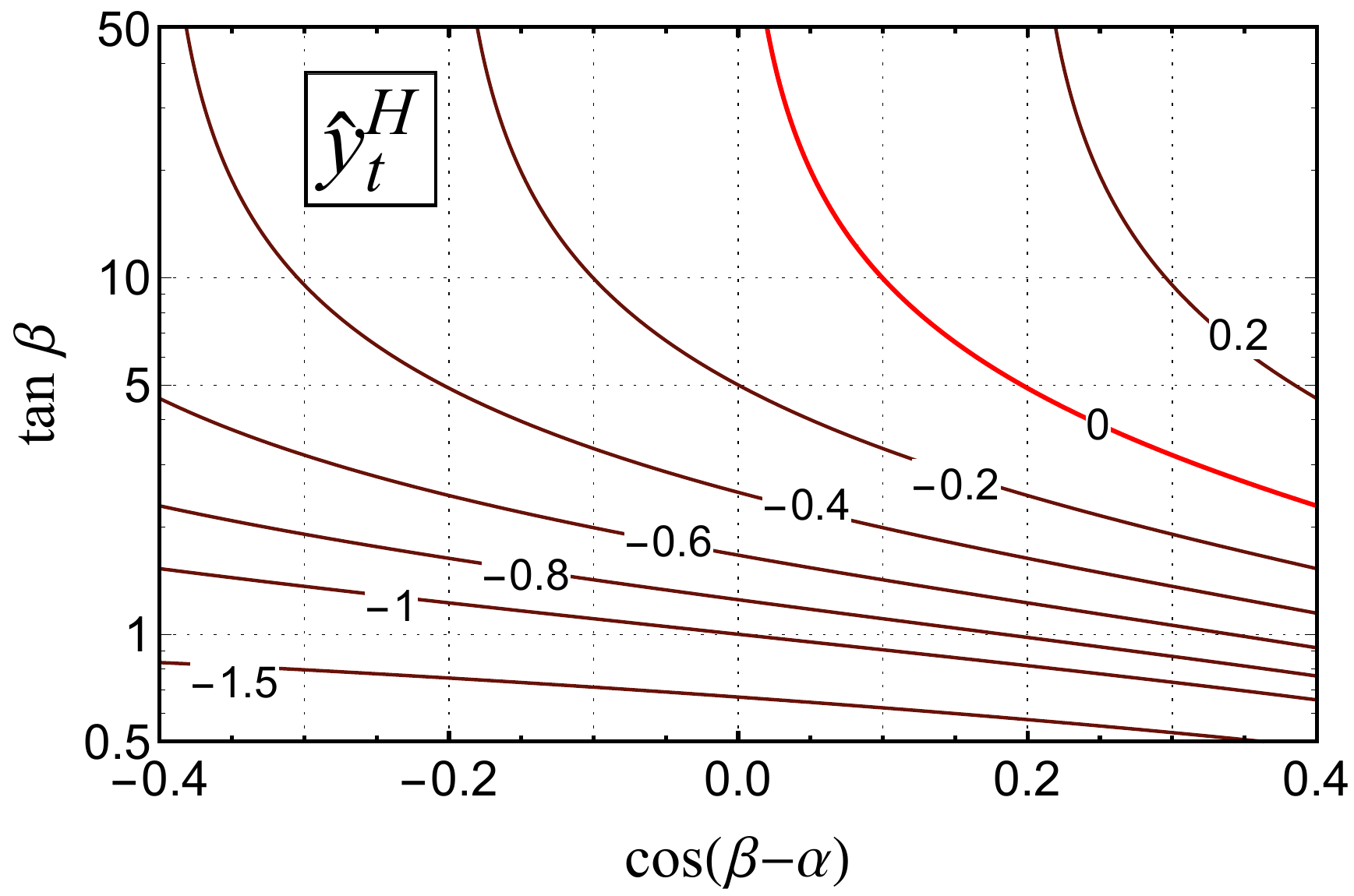}
\vspace*{-0.2cm}
\caption{\label{fig:UpYukawa}
\baselineskip 3.5ex
The top quark Yukawa coupling $\yh^H_{t}$ normalized by the SM one is shown in the $(\cba,\tb)$ plane
for $\sba>0$. We mark the top-phobic $\yh^H_t =0$ contour with brighter red.
}
\vspace{-0.2cm}
\end{figure}

We find that there exist a special parameter choice for $\yh^H_{t} =0$,
called the top-phobic $H^0$.
To be more specific, we present the value of $\yh^H_{t}$ in the parameter space
of $(\cba,\tb)$ in Fig.~\ref{fig:UpYukawa}.
As the red line indicates,
a specific \emph{nonzero positive} $\cba$ for a given $\tb$
leads to vanishing $\yh^H_{t}$,
which happens,
for example, when $\tb=10$ for $\cba=0.1$ or
$\tb=2.3$ for $\cba=0.4$. Near the top-phobic line the signal rate is severely suppressed especially for Type I.

\subsection{General formalism for interference}
We consider the interference between the
continuum background and the resonance process
of a particle with mass $M$ and total decay width $\Gamma$ in a $2 \to 2$ scattering process.
When we write the helicity amplitudes for the continuum background ($\mathcal{M}_{\rm cont}$) and the resonance ($\mathcal{M}_{\rm res}$) as
\bea
\label{eq:amp:def}
\mathcal{M}_{\rm cont} &=&{\calA}_{\rm cont} e^{i \phi_{\rm cont}}, \\ \no
\mathcal{M}_{\rm res} &=& \frac{M^2}{{\hat s}-M^2+iM\Gamma}\; {\calA}_{\rm res}e^{i \phi_{\rm res}},
\eea
the total partonic cross-section becomes
\begin{eqnarray}
\label{eq:formalizm}
{\hat\sigma}_{\rm cont} +   {\hat \sigma}_{\rm sig}  
&=&  {\hat\sigma}_{\rm cont} +   {\hat \sigma}_{\rm res}
\frac{M^4}{(\shat-M^2)^2+M^4w^2}
\left[ 1+  \frac{2w}{R}\sphi +\frac{2(\shat-M^2)}{M^2}\frac{ \cphi}{R} \right]
\\ \no
&\equiv&
{\hat\sigma}_{\rm cont} +   {\hat \sigma}_{\rm res}
\Big[
f_{\rm BW}(m_{\rm inv}) + f_{\rm Im}(m_{\rm inv}) + f_{\rm Re}(m_{\rm inv})
\Big],
\end{eqnarray}
where $m_{\rm inv}=\sqrt{\hat{s}}$. Note that $f_{\rm BW}, f_{\rm Im}, f_{\rm Re}$ take the terms in square bracket one by one.
${\hat \sigma}_{\rm cont}$, ${\hat \sigma}_{\rm res}$, $R$,
and the interference phase $\phi$ are
\bea\label{eq:phi:definition}
{\hat \sigma}_{\rm cont, res} &=&  \frac{1}{32\pi\shat}
 \int dz \sum \calA_{\rm cont, res}^2,
 \\
\nn
{\hat \sigma}_{\rm int} e^{i\phi} &=&
\frac{1}{32\pi\shat} \int dz \sum \calA_{\rm cont}\calA_{\rm res} e^{i(\phi_{\rm res}
- \phi_{\rm cont})}, \, \\ \nn
R &=& \frac{{\hat \sigma}_{\rm res}}{{\hat \sigma}_{\rm int}}~ ,
\quad w\equiv \frac{\Gamma}{M},
 \eea
where $z = \cos\theta^*$ while
$\theta^*$ is the scattering angle in the c.m.~frame.
The summation is over helicity
and color indices.
$R$, $w$, and $\phi$ are the key parameters
which determine the pattern of interference effect.
More intuitive form for $R$ and $\phi$ can be obtained if
assuming that one helicity amplitude is dominant:
\begin{equation}
\label{eq:approximate:R:phi}
R\simeq \frac{\calA_{\rm res}}{\calA_{\rm cont}},
\quad
\phi\simeq \phi_{\rm res}-\phi_{\rm cont}\,.
\end{equation}
As can be understood from Eq.~(\ref{eq:formalizm}) and will be discussed more, $w/R$ indicates the strength of interference effect and $\phi$ determines whether it is imaginary-part interference ($c_\phi = 0$) or real-part interference ($s_\phi = 0$), or between the two.

Most of new particles of our interest have narrow width ($w \ll 1$),
which confines the signal events in the resonance region
of the invariant mass  distribution.
It is a good approximation to ignore the $m_{\rm inv}$ dependence of $R$ and $\phi$.
Then $m_{\rm inv}$ dependence of ${\hat \sigma}_{\rm sig}$ is explicitly shown in  Eq.~(\ref{eq:formalizm}) as a simple function of $\hat s$ ($=m_{\rm inv}^2$).
Apparently,
$f_{\rm Re}(m_{\rm inv})$ is an odd function at $m_{\rm inv}=M$,
which yields a dip-peak or peak-dip structure.
On the contrary $f_{\rm BW}(m_{\rm inv})$ and  $f_{\rm Im}(m_{\rm inv})$
are even functions.
The sensitivity to $f_{\rm Re}(m_{\rm inv})$ and  $f_{\rm Im}(m_{\rm inv})$
crucially depends on the bin size of the invariant mass distribution.
If the bin is large such that a dip-peak structure is included in one bin,
we should integrate over $m_{\rm inv}$,
which eliminates the real-part interference.
If the bin is narrow enough,
more dynamic structure of  $f_{\rm Re}$ can be probed.
We consider these two cases and suggest new observation factors for each case.

\textit{(i) Large bin:}
In this case, we integrate $\hat{\sigma}_{\rm sig}$ over $m_{\rm inv}$,
under which the even functions survive but the odd function $f_{\rm Re}(m_{\rm inv})$ is washed out
at leading order\footnote{Of course, the cancelation is not perfect
because of the strong $\sh$ dependence of the gluon luminosity.}.
The survived imaginary part interference results in a multiplicative factor,
$(1+2 w \sphi/R)$, to the NWA rate $\sigma\cdot\br$. Therefore, the total signal rate can significantly differ from what obtained from the NWA due to the imaginary part interference.
In Ref.~\cite{Jung:2015gta}, we called this as the correction factor $C(\equiv 1+\Delta C)$:
\bea
\label{eq:correction}
C \equiv \frac{\sigma_{\rm mNWA}}{\sigma_{\rm prod} \cdot \br}=
1+\frac{2w}{R} \sphi
\,,
\eea
where $\sigma_{\rm mNWA}$, whose subscript denotes {\it modified} NWA, is the total signal rate by including imaginary-part interference effect.
In the pure imaginary case ($\cphi = 0$)
there are three unique shapes of a resonance:
a pure dip ($C<0$),
a pure peak ($C>0$),
or a nothingness ($C=0$).
Note that the $C$ factor is measurable by comparing the
observed event rate with the simulation result of $\sigma_{\rm prod}\cdot\br$.

\begin{figure}[t]
\includegraphics[width=0.52\textwidth]{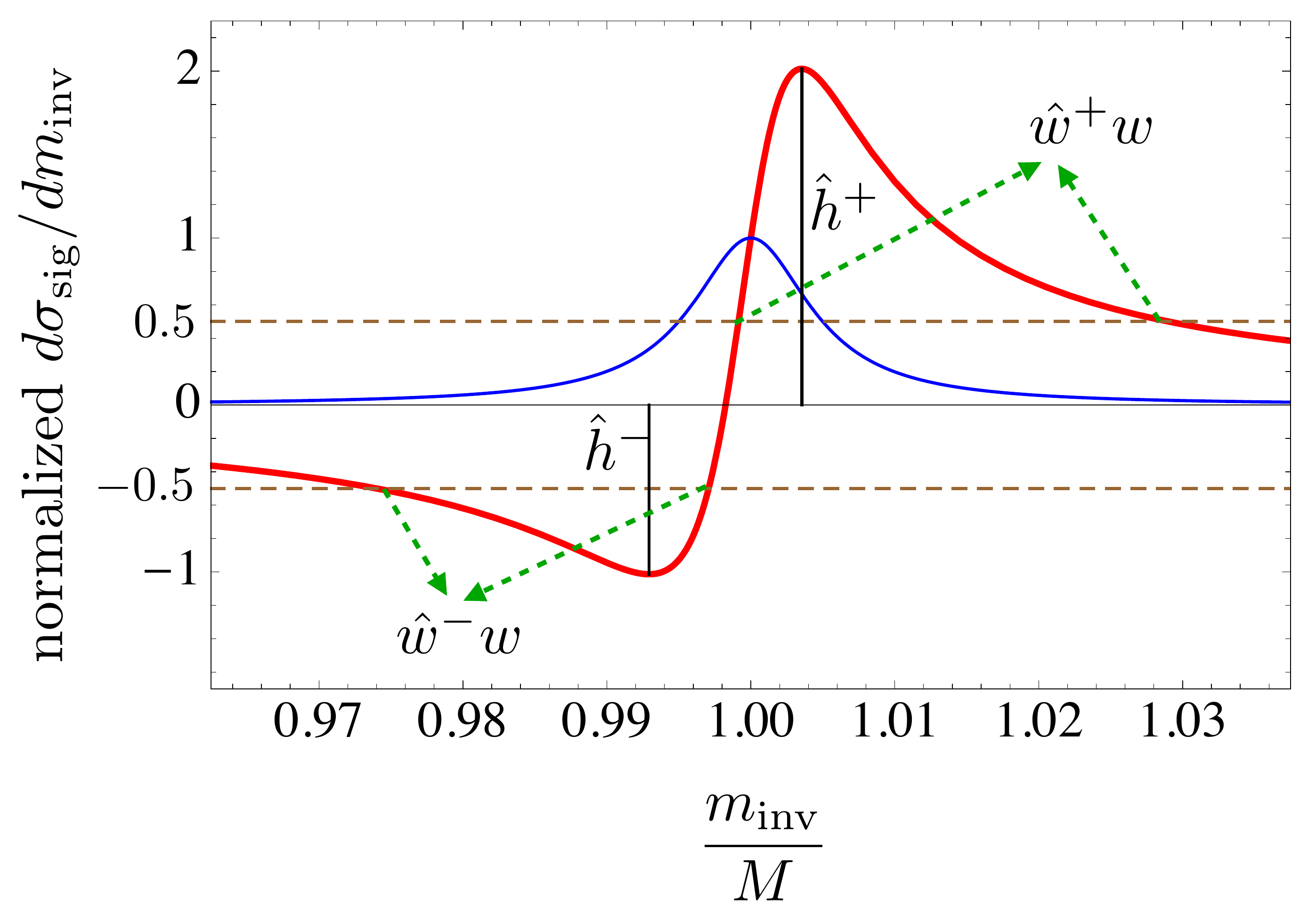}
  \vspace*{-0.2cm}
  \caption{\label{fig:D:definition}
\baselineskip 3.5ex
The definitions of variables characterizing a resonance shape in Eq.~(\ref{eq:hpm:wpm:general}): the relative height $\hat{h}^+$ and depth $\hat{h}^-$, the relative peak width $\hat{w}^+$,
and the relative dip width $\hat{w}^-$
for $w = 0.01$, $R = 0.007$, and $\phi = 0$.
We normalized the differential cross section about the invariant mass
such that the peak height of the BW resonance without any interference (blue)
is a unity.
}
  \vspace{-0.2cm}
\end{figure}

\textit{(ii) Fine bin:}
In an ideal situation with very small bin size,
the $m_{\rm inv}$ dependence of $\sigma_{\rm sig}$
can be measured, which is more crucial in observing
the dip-peak structure with nonzero real-part interference.
When $\cphi \neq 0$,
the full resonance shape of $\hat{\sigma}_{\rm sig}(m_{\rm inv})$
is a dip-peak (peak-dip) structure if ${\rm sign}(\cphi)=+1(-1)$.
In order to quantify the signal rates of the dip and peak,
we define the relative height $\hat{h}^+$ and depth $\hat{h}^-$,
compared with the height of the BW peak.
In addition, the relative widths of the peak and dip are
defined as $\hat{w}^\pm$, as illustrated in Fig.~\ref{fig:D:definition}.
For the general case
with both real- and imaginary-part interference,
we obtain $\hat{h}^\pm$ and $\hat{w}^\pm$ in terms of our key parameters as
\bea
\label{eq:hpm:wpm:general}
\hat{h}^\pm &=& \frac{1}{2}
\left[
\sqrt{C^2+\left( \frac{2 w \cphi}{R} \right)^2 }
\pm C
\right], \\ \no
\hat{w}^\pm &=&
\sqrt{\left( \frac{2 w \cphi}{R} \right)^2 -1 \pm2 C}.
\eea
Note that $\hat{w}^-$ is not defined if $\left({2 w \cphi}/{R} \right)^2 <1 +2 C$ when $C>0$:
this is because we defined $\hat{w}^-$ as the width at the
negative half-maximum of the BW one.
If the dip is shallow such that its depth is smaller than the BW half-maximum,
we do not calculate $\hat{w}^-$ as considering it as a shallow dip.

Two limiting cases ($2w/R \ll 1$ and $2w/R \gg 1$)
present a clear understanding of the real part interference effect.
When $\sphi=0$, Eq.~(\ref{eq:hpm:wpm:general}) to leading order becomes
\bea
\label{eq:large:wR}
\hat{h}^+  \simeq \hat{h}^- \simeq\frac{w}{R},
\quad ~~\hat{w}^+\simeq\hat{w}^+\simeq \frac{2w}{R}
,&& ~~ \quad \hbox{if } \frac{2w}{R}\gg 1;
\\ \label{eq:small:wR}
\hat{h}^+ \simeq 1
, \quad
\hat{h}^- \simeq 0
,
\quad \hat{w}^+ \simeq 1,&& ~~ \quad
\hbox{if } \frac{2w}{R}\ll 1.
\eea
It is clear that the total width
is not the
key parameter which determines the real part interference, but the ratio  $w/R$ is.
Even though $\Gamma$ is very small compared to its mass $M$,
smaller $R$
can make the ratio $w/R$ large:
we have a profound dip-peak structure with enhanced height and enhanced width:
see Eq.~(\ref{eq:large:wR}).
If $w/R \ll 1$, we have a very shallow dip and a BW-like peak as shown in Eq.~(\ref{eq:small:wR}).

On the analogy of the correction factor $C$,
which quantifies the integrated NP signal rate with the interference,
we suggest two new factors, the distribution factors $D^+$ and $D^-$,
defined by
\bea
D^\pm = \hat{h}^\pm \hat{w}^\pm.
\label{eq:dpm}
\eea
Note that $D^\pm$ is not exactly proportional to the new physics signal rate,
since the resonance shape with interference is different from the ordinary BW form.
However these simple factors provide a powerful estimate for the dip-peak structure,
especially useful when scanning a theoretical parameter space
in order to look for large real-part interference effect.
If $D^- =1$ for example,
we can expect that a dip shall appear with almost the same rate with the BW resonance.

\section{$\gamma\gamma$ channel}
\label{sec:rr}
The diphoton decay channel of a heavy neutral Higgs boson $H^0$ produced by the gluon-gluon fusion is
a two loop process while
the SM continuum background $gg\to\rr$
is a one loop process\footnote{
\baselineskip 3.5ex
We do not consider subdominant two-loop contribution in continuum background. Its effect on the resonance-continuum interference was shown to be
$5\%$ when $M_H < 160\GeV$~\cite{Dixon:2003yb,Bern:2001df}.}.
%
The parton level differential cross section
of $gg\to \gamma\gamma$
is \begin{equation}
\frac{d\sgh (gg\to \rr)}{dz} = \frac{1}{32\pi \hat s} \frac{\alpha_s^2 \alpha_e^2}{2}  \sum_{\lambda_1 \lambda_2 \lambda_3 \lambda_4}
\left| {\cal M}^{\rm cont}_{\lambda_1 \lambda_2 \lambda_3 \lambda_4}
+
{\cal M}^H_{\lambda_1 \lambda_2 \lambda_3 \lambda_4} \right|^2\,,
\end{equation}
where ${\cal M}^{{\rm cont},H}_{\lambda_1 \lambda_2 \lambda_3 \lambda_4}$
are the normalized helicity amplitudes
from the continuum background and $H^0$ resonance with
the helicity of incoming gluons ($\lambda_1, \lambda_2$) and outgoing photons ($\lambda_3, \lambda_4$).
The overall factor is due to the loop correction.
We refer the explicit expressions of ${\cal M}^{\rm cont}$'s to
Refs.~\cite{Constantini:1971,Dicus:1987fk}.
For the Higgs resonance signal $gg \to H \to \rr$,
only four helicity amplitudes
are nonzero:
\begin{eqnarray}
\label{eq:ampH}
{\cal M}_{++++}^{H} &=& {\cal M}_{----}^{H} =  {\cal M}_{++--}^{H}
=  {\cal M}_{--++}^{H} \nn \\
& =& \frac{ G_F}{128 \pi^2} \frac{\hat s^2}{\hat s-M_H^2 + i M_H \Gamma_H}
\sum_q {\hat y}_q^H A^H_{1/2}\left(\tau_{q}\right) \nn \\
&& \times \bigg( \sum_q {\hat y}_q^H N_c Q_q^2 A^H_{1/2}\left(\tau_{q}\right)
+ \sum_\ell {\hat y}_\ell^H Q_\ell^2 A^H_{1/2}\left(\tau_{\ell} \right)
+ \cosbma A^H_1(\tau_W)
\bigg) \,,
\end{eqnarray}
where $\tau_p = \hat{s}/(4m_p^2)$, and $\yh^H_{q,\ell}$'s are shown in Eq.~(\ref{eq:Yukawa:couplings}),
and the expressions of $A_{1/2,\,1}^H(\tau)$
are in Ref.~\cite{Djouadi:2005gj}.

We first study the relative complex phase $\phi$.
The complex phase arises through the loop diagrams when
the squared of the momentum that passes through an internal cut line is greater than
the threshold mass square in the loop.
The continuum background $gg\to \rr$ is dominated by
light quark ($u,d,s,c,b$) loops whose complex phase arises in $(+-+-), (-+-+), (+--+), (-++-)$ helicity amplitudes. But, those helicity amplitudes do not inference with the Higgs resonance amplitudes as can be clearly seen
in Eq.~(\ref{eq:ampH}). Then, only the top quark loop can give small contribution to the complex phase.
The relative phase $\phi$
is mainly from the Higgs resonance.
$gg\to H \to \rr$ is dominated by top quark loop and $W$ boson loop
where the latter is suppressed for small  $\cosbma$.
When $M_H < 2m_t$ the imaginary part of amplitude arises mostly from the $W$ loop while
 after $M_H > 2m_t$ from the top quark loop.
 If both $\cba$ and ${\hat y}_t^H$ are positive like the SM Higgs boson,
 two contributions
 are destructive.

\begin{figure}[t]
  \centering
  \includegraphics[width=0.96\textwidth]{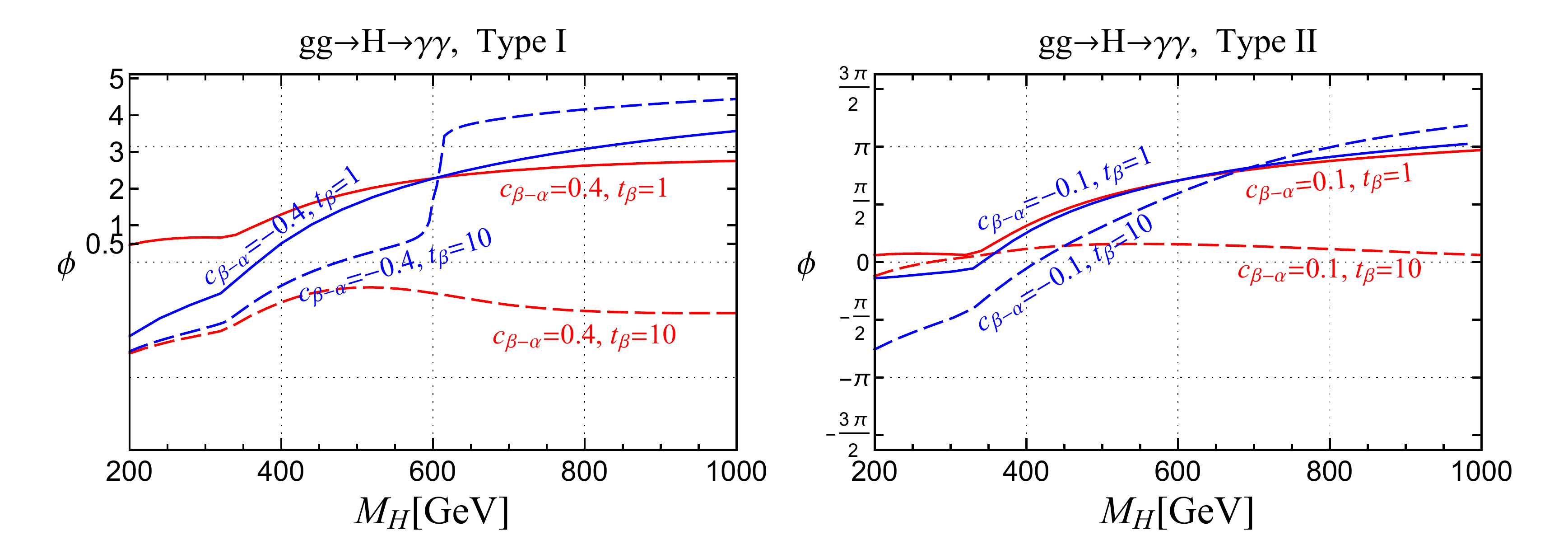}
  \vspace*{-0.2cm}
  \caption{\baselineskip 3.5ex
  The resonance-continuum interference phase $\phi$ of the $gg\to (H\to) \rr$ process for various benchmark parameter values of $\cosbma$ and $\tb$ in the Type I (left panel) and Type II (right panel) models. The abrupt change at around $M_H \simeq 2m_t$ is due to the onset of top-pair threshold.
  }
  \label{fig:phirr}
  \vspace{-0.2cm}
\end{figure}

In Fig.~\ref{fig:phirr}, we show  $\phi$ with respect to $M_H$ for several benchmark parameter points of $\cosbma$ and $\tb$.
We set $\cosbma=\pm 0.4$ for Type I
and $\cosbma=\pm 0.1$ for Type II,
which are marginally allowed by the current Higgs precision measurement.
In both Types, there is a considerable portion of parameter space
where the imaginary-part interference is large, \ie sizable $\sphi$.

For Type I, we consider four cases of $\cosbma=\pm 0.4$
and $\tb=1,10$.
The value of $\phi$
is determined by the behaviors of the real and imaginary parts
of the loop functions at the given mass $M_H$.
For example, let us focus on $M_H=200\gev$ where $W$ loop mainly generates a phase.
The case $(\cba=0.4,\tb=1)$ has $\phi$ near zero
while the other three cases have large negative values near $-\pi/2$.
As shown in Fig.~\ref{fig:UpYukawa},
$(\cba=0.4,\tb=1)$ leads to negative ${\hat y}_t$
so that the top quark loop and the $W$ loop become constructive:
the real part of the total loop function become larger;
the phase $\phi$ is small.
For the other three cases, the destructive interference
reduces the real parts of the loop function significantly.
The untouched imaginary loop function part, which is negative in these cases,
yields $\phi \sim -\pi/2$.
For other $M_H$,
similar understanding is possible through the relative
strengths and signs of $\yh_{t,b}$ and $\cba$.
The abrupt change of amount $\pi$ in $\phi$ at $M_H=600\gev$ for $\cba=-0.4$ and $\tb=10$ is attributed to the cancelation of two real-parts between $W$ and top loops and consequent sign flip.

For Type II, the four cases of $\cba=\pm0.1$ and $\tb=1,10$ are considered.
The small values of $\cba = \pm 0.1$ suppress
the $W$ loop contribution.
The imaginary phase occurs in the loop function $A^H_{1/2}(\tau_t)$
when $M_H> 2 m_t$.
For $M_H<2m_t$,
$\phi$ is close to zero  except for
the case $(\cosbma=-0.1, \tb = 10)$
where the sizable $b$ quark contribution with large $\tb$
considerably cancels the whole real part
and $\phi$ has large negative value.
As $M_H$ crosses the $2m_t$ threshold,
$\phi$ continually increases up to $\pi$
except for $(\cosbma=0.1, \tb = 10)$ case.
Here the real part of $A^H_{1/2}(\tau_t)$ vanishes when $M_H=1.1\tev$
so that one loop function generates the phase of $\pi/2$.
Two factors of $A^H_{1/2}(\tau_t)$
from production and decay processes yields $\phi=\pi$.
The exceptional $(\cosbma=0.1, \tb = 10)$ case is
near the top-phobic line.
In this case, real parts of $W$ loop and bottom quark loop are added up while some cancelation arise for imaginary part. Therefore, $\phi$  is very small along the whole $M_H$ region.

Secondly we examine
$w$ and $R$ in $gg\to \rr$,
of which the ratio $w/R$ is the crucial factor to determine
the interference effects.
For simplicity, we assume that $H^0$ decays
into $WW,ZZ,q\bar{q},\ttau,\rr$.
The possibly important decay channel $H\to hh$ is neglected,
which can be achieved by tuning the soft $Z_2$
symmetry breaking term $m_{12}^2$.
Since this assumption minimizes the total width and thus
the interference effects,
it shows a good limiting feature of the interference effects in $gg\to \rr$.
The effect of additional decay channels shall be discussed in Sec.~\ref{sec:extra}.
Under this assumption,
the $\Gm_H$, which generally increases with $M_H$, depends on  $\cba$ and $\tb$.
If $\cosbma=\pm 0.1$, the parameter $w$ for $M_H\lsim 1\tev$ is
very small to be ${\cal O}(10^{-3}) - {\cal O}(10^{-2})$
depending on the $\tb$ value
since $H \to WW,ZZ$ is very suppressed.
If $\cosbma=\pm 0.4$ the $WW,ZZ$ decay channels becomes quite significant
so that $\Gm_H$ can be $100\GeV$ until $M_H<1\TeV$.

The $R$ value, proportional to
the ratio of the magnitude of the signal amplitude
to that of the background amplitude,
is generically small in $gg\to\rr$.
The background process is at one loop
and the $H^0$ signal is at two loop.
The ratio $R$ is roughly one-loop suppressed.
In most of the parameter space $R$ is less than $10^{-3}$.
We have very large interference effects.

\begin{figure}[t]
  \centering
 \includegraphics[width=0.96\textwidth]{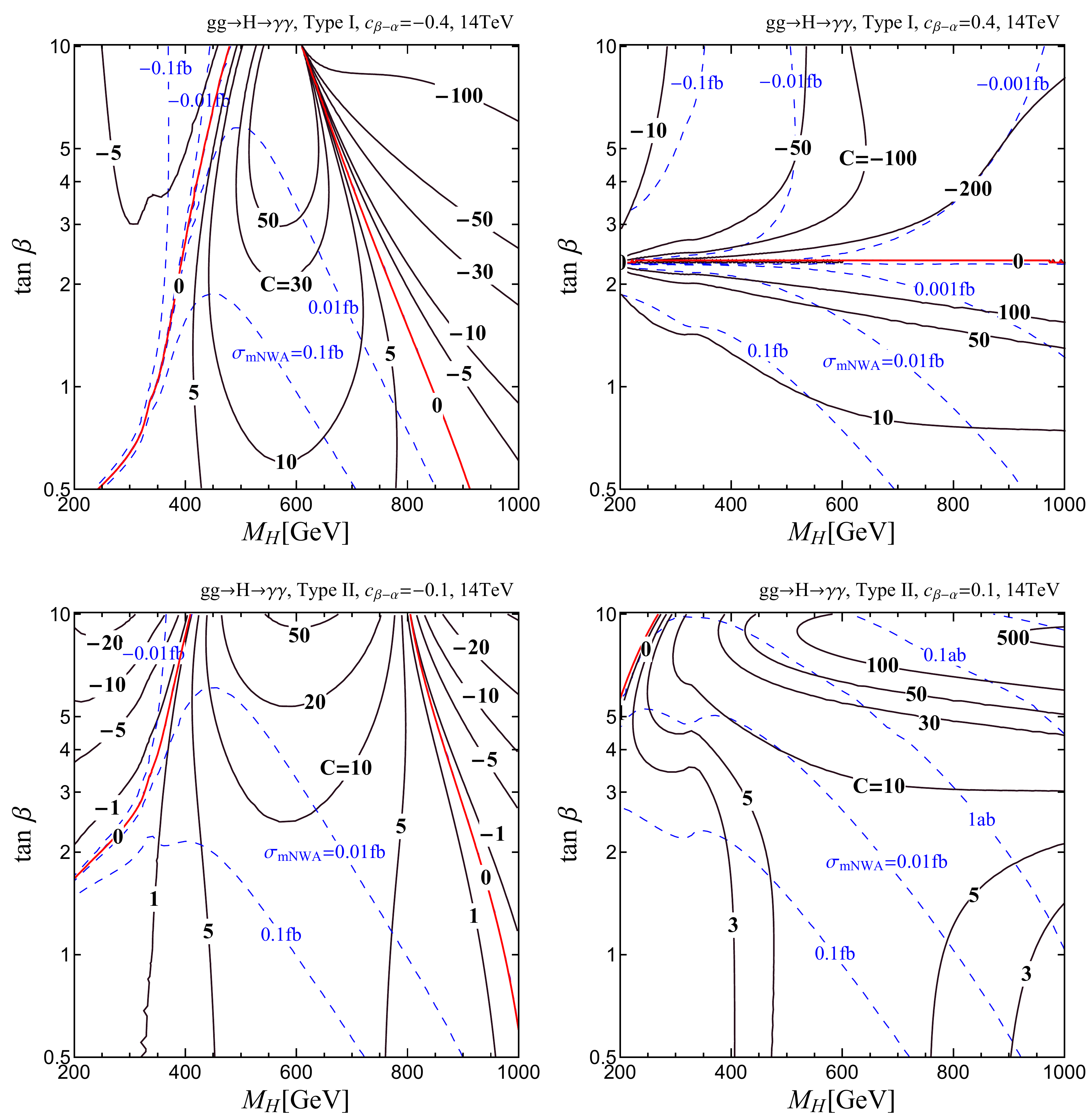}
 \vspace*{-0.2cm}
  \caption{\baselineskip 3.5ex
  Contours of the $C$ factor (solid black) and the
  $\sigma_{\rm mNWA}$ (dashed blue) of the $gg\to H\to \rr$ process at the LHC $14\TeV$ in the ($M_H$, $\tb$) plane. The contour of $C=0$ is marked with red-solid.
   }
  \label{fig:Crr}
  \vspace{-0.2cm}
\end{figure}

Now we quantitatively discuss the interference effects.
In Fig.~\ref{fig:Crr}, we show the contour plots for $C$ factor as well as
the modified total signal rate $\sg_{\rm mNWA} = C \times(\sg\cdot \br)$ at $14\TeV$ in $(M_H, \tb)$ plane
for $\cba=\pm0.4$ in Type I and $\cba=\pm 0.1$ in Type II.
For the initial gluon luminosity, we used CT10NLO PDF set~\cite{Lai:2010vv}.
We also applied NNLO $k$-factor to the heavy Higgs resonance production part using \texttt{HIGLU} fortran package~\cite{Spira:1995mt} and use the LO decay rate of heavy Higgs
in Ref.~\cite{Song:2014lua}.

The most unexpected result is
that the interference effect $|\Delta C|$ can be ${\cal O}(100)\%$
even when $w$ is sub-percent level ($\cosbma=\pm 0.1$ and large $\tb$).
The usually adopted criteria to ignore the interference effect,
$w\ll 1$, is not good enough.
The characteristics of $C$ values for different type and parameters
are as follows.
The $C$ contours for $\cosbma < 0 $
show some common features in both Type I and Type II.
They have three sectors
divided by two $C=0$ lines.
The left and right parts of these lines
have large negative $C$ factors (dip-like resonance shapes), while the middle region has
positive $C$ factors (peak-like resonance shapes).
The two $C=0$ lines appear near two points $\phi\simeq 0$ and $\phi\simeq\pi$,
where $\Delta C$ flips its sign
(here $2w/R$ is very large).
As shown in Fig.~\ref{fig:phirr}(b),
crossing $\phi=0,\pi$ happens two times for $200\gev <M_H<1\tev$,
which passes near two $C=0$ lines.
In addition, the magnitude of $C$ increases with increasing $\tb$.
This is because the signal amplitude (or $R$) is reduced by increasing $\tb$.
On the contrary $\sg_{\rm mNWA}$ decreases with increasing $\tb$.
In the Type II $\cba=0.1$ case, there is only one $C=0$ line since $\phi$ crosses $\phi=0$ point
once when $\tb=10$: see Fig.~\ref{fig:phirr}(b).


For the $\cba>0$ case of Type I,
there is one horizontal $C=0$ line.
However the origin of $C=0$ here is
different from the other three cases.
It
is very close to the top-phobic (actually fermion-phobic
due to the common $\yh^H$'s for all fermions) line
as in Fig.~\ref{fig:UpYukawa}.
This fermion-phobic nature prohibits the gluon fusion production itself.
In the vicinity of the $C=0$ line, very small $R$ is generated,
enhancing $C$ extremely.
Since the sign of top Yukawa coupling is flipped at this line, so does the sign of $C$.
Another unexpected result is that the large $\tb$ region in Type I
has compatible signal rate with the small $\tb$ region
since $\yh_t^H$ increases again with $\tb$ after crossing the top-phobic line.
Our final observation is that
$|\dtc|$ and $\sigma_{\rm mNWA}$
are anti-correlated in general:
the region for large $|\dtc|$ usually has very suppressed signal rate.

\section{ZZ channel}
\label{sec:ZZ}
 For the $ZZ$ channel, both the SM background process $gg\to ZZ$
 and the Higgs resonance signal $gg\to H \to ZZ$ are one-loop processes.
The partonic differential cross section
 is
\begin{equation}
\label{eq:gg2zz}
\frac{d \hat \sigma (gg\to ZZ)}{dz} = \frac{1}{32\pi \hat s}\frac{\alpha_s^2\alpha_Z^2\beta_Z}{256}\sum_{\lambda_1\lambda_2\lambda_3\lambda_4}
|{\cal T}^{\rm cont}_{\lambda_1\lambda_2\lambda_3\lambda_4}
+{\cal T}^{H}_{\lambda_1\lambda_2\lambda_3\lambda_4}|^2\,,
\end{equation}
where $\alpha_Z = \alpha_e /(\sin^2 \theta_W \cos^2 \theta_W)$ with weak mixing angle $\theta_W$ and $\beta_Z=\sqrt{1-4m_Z^2/{\hat s}}$.
${\cal T}^{\rm cont}$  and ${\cal T}^{H}$
are the normalized helicity amplitude
for the background and $H^0$ signal, respectively,
of which the expressions are referred to Ref.~\cite{Glover:1988rg}\footnote{
\baselineskip 3.5ex
${\cal T}_{\lambda_1\lambda_2\lambda_3\lambda_4}$ are corresponding to ${\cal M}_{\lambda_1\lambda_2\lambda_3\lambda_4}$ in Ref.~\cite{Glover:1988rg}. There is an obvious typo in Eq.~(3.14) of  Ref.~\cite{Glover:1988rg} since the mass dimensions of the first two terms in curly brackets of right hand side are incorrect. From the independent calculation, we find that it should be $(1+\beta)\{4s(t-u)Y/(s_4 u_1 t_1) + 4 Y /s_4 (...)+...\}$ where the ellipsis represent the same form of the equation.
}.
A big difference between ${\cal T}^{\rm cont}(gg\to ZZ)$
and ${\cal M}^{\rm cont}(gg\to\rr)$ is the presence of the longitudinal mode in the $ZZ$ channel.
In particular when both outgoing $Z$ bosons are longitudinal
(called the LL mode),
the amplitude is proportional to $m_q^2$.
In the $ZZ$ mode, therefore,
the top quark contribution becomes important
after the $\ttbar$ threshold.
We found that the LL contribution
increases linearly with $m_{ZZ}$,
reaching about $50\%$ at $m_{ZZ}=700\GeV$.

\begin{figure}[t]
  \centering
  \vspace*{-0.2cm}
 \includegraphics[width=0.96\textwidth]{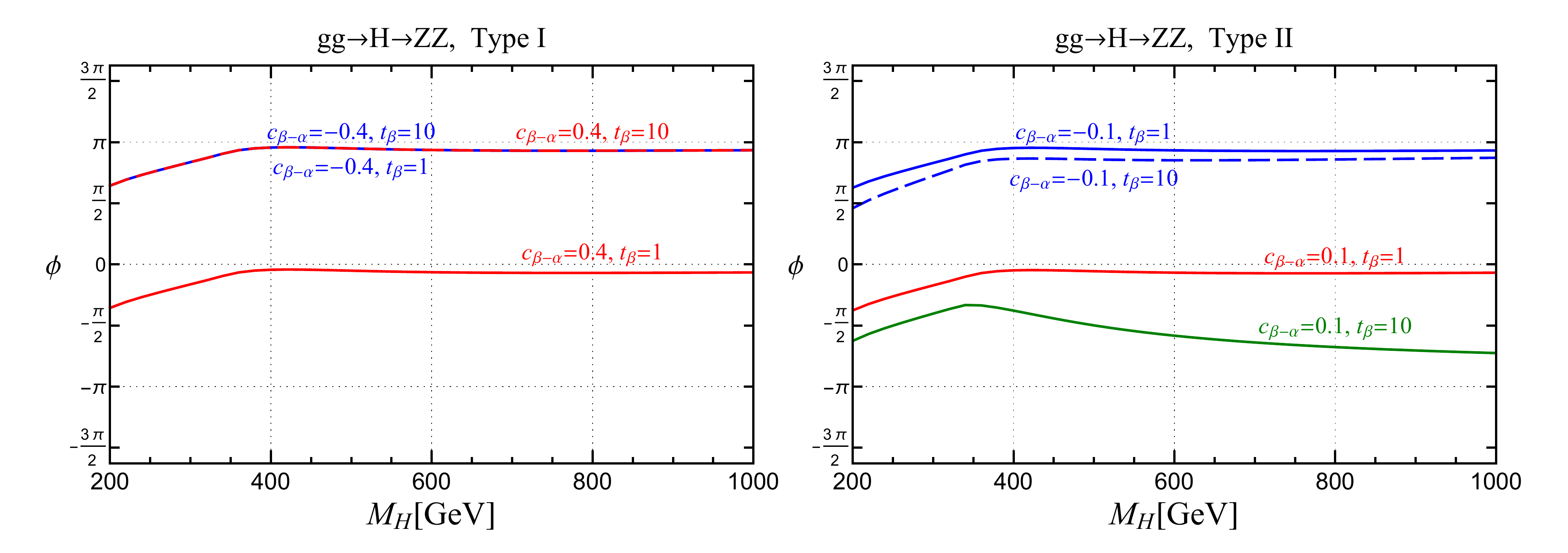}
  \caption{\baselineskip 3.5ex
    The resonance-continuum interference phase $\phi$ of the $gg\to (H\to) ZZ$ process for various benchmark parameter values of $\cosbma$ and $\tb$ in the Type I (left panel) and Type II (right panel) models. The abrupt change at around $M_H \simeq 2m_t$ is due to the onset of top-pair threshold. The top-phobic case is shown as the green-solid.
  }
  \label{fig:phiZZ}
  \vspace{-0.2cm}
\end{figure}

We first study the complex phase $\phi$ in $ZZ$ channel.
For the background process,
the complex phase from the top quark loop significantly
increases with invariant mass $m_{ZZ}$ after the $\ttbar$ threshold
while the light quark contribution to $\phi_{\rm cont}$ decreases quickly.
For $gg \to H \to ZZ$,
the top quark loop contribution to $gg\to H$
provides dominant complex phase unless $b$ contribution becomes dominant
for large $\tb$ in Type II.
It appears that the behaviors of complex phases of
the top loop contributions from $gg \to ZZ$ and $gg \to H \to ZZ$ are quite similar after $M_H > 2m_t$.
As a result, the phase $\phi\sim(\phi_{\rm res}-\phi_{\rm cont})$
is close to $0$ or $\pi$ depending on overall sign.
This is clearly shown in Fig.~\ref{fig:phiZZ}.
One exception is the case $(\cosbma=0.1, \tb=10)$ in Type II (green-colored line),
which corresponds to top-phobic $H^0$:
$\phi$ is quite different from 0 or $\pi$.
Here the bottom quark loop is dominant.
We also found that Type I shows special behavior: only two curves appear
regardless of $\cosbma$ and $\tb$.
It is because all $\yh^H$'s are the same and thus
 $\phi_{\rm res}$ is also the same except the overall sign.
 For example, $\cba \yh^H$ in the case $(\cosbma=0.4, \tb=1)$
  has opposite sign of
  that in the other three cases: see Fig.~\ref{fig:UpYukawa}.

\begin{figure}[t]
  \centering
    \vspace*{-0.2cm}
 \includegraphics[width=0.96\textwidth]{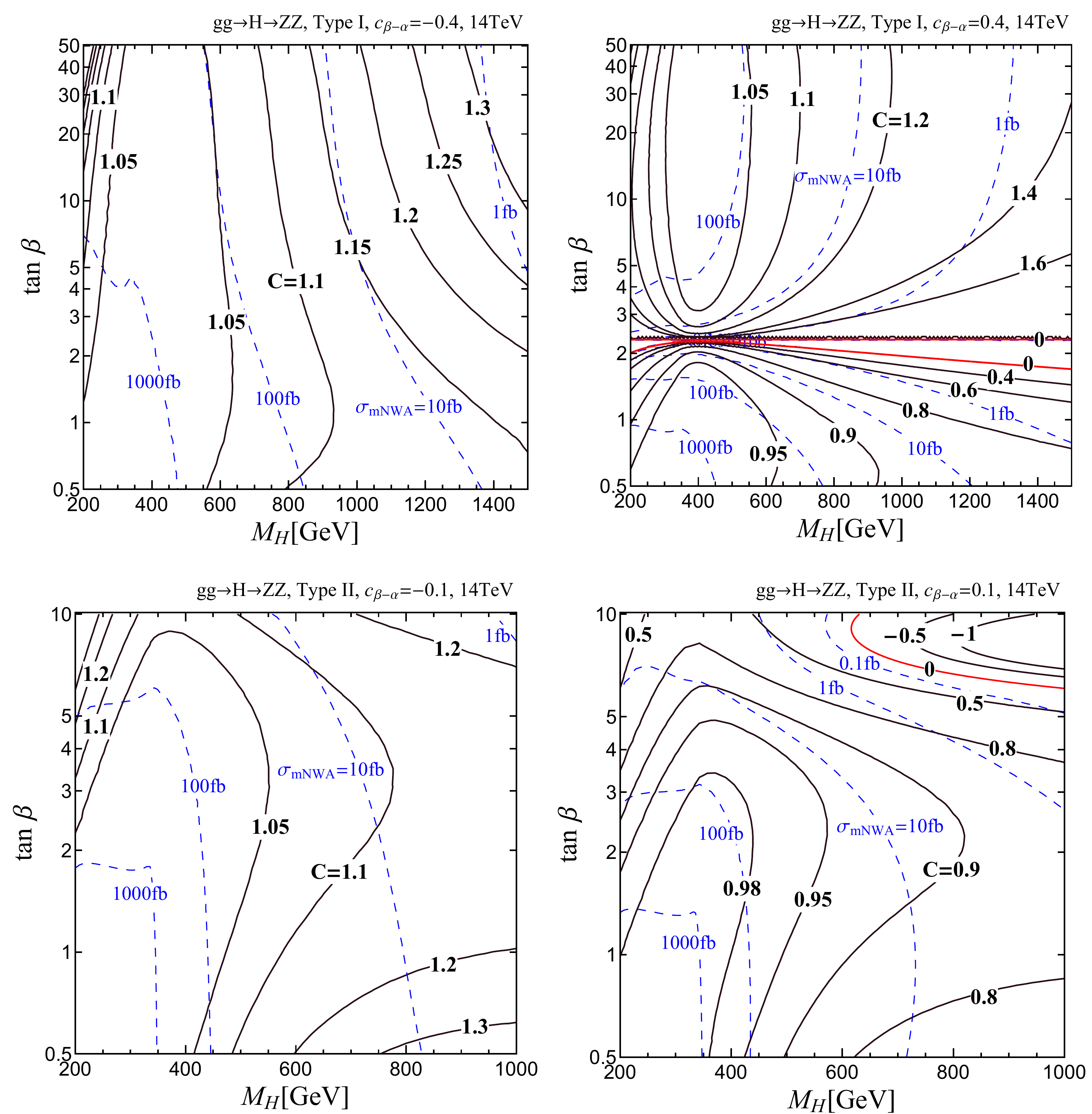}
  \caption{\baselineskip 3.5ex
    Contours of the $C$ factor (solid black) and the
  $\sigma_{\rm mNWA}$ (dashed blue) of the $gg\to H\to ZZ$ process at the LHC $14\TeV$ in the ($M_H$, $\tb$) plane. The contour of $C=0$ is marked with red-solid.
   }
  \label{fig:CZZ}
    \vspace{-0.2cm}
\end{figure}

Now, we investigate $C$ factor and $\sigma_{\rm mNWA}$.
Basic setup is similar to $\rr$ channel.
The $w$ parameter is small in general.
The $R$ is roughly ${\cal O}(0.01)$, mostly larger than $w$,
since
both the SM background process and the Higgs resonance signal are one-loop processes.
The interference effect would be generically small in $ZZ$ channel for small $w$.
Moreover the $\cosbma$ value does not affect the ratio $w/R$ much
since larger $\cba$ increases both $R$
and  $w$.

In Fig.~\ref{fig:CZZ}, we show the contours of the $C$ factor
and $\sigma_{\rm mNWA}$ in $(M_H, \tb)$ plane for $\cba=\pm0.4$ for Type I
and $\cba=\pm0.1$ for Type II.
The first important result is that
the interference effect is not negligible even when $w$ is very small,
though not dramatic as
in the $\rr$ channel.
For example, the case of $M_H=300\GeV$, $\cosbma=0.1$, and $\tb=10$
in Type II, where $\Gamma_H/M_H = 0.2\%$, has $\dtc \simeq -30\%$.
The second result is
that for the given $\tb$, $|\Delta C|$ decreases with $M_H$
but increases again after the $\ttbar$ threshold.
Before the $\ttbar$ threshold $R$ increases much faster than $w$.
After the $\ttbar$ threshold,
$w$ increases more sharply
as the $\ttbar$ decay channel is opened.

The negative $\cba$ case has very smooth and moderate variation of $C$.
We have maximally 30\% of $\dtc$ for large $\tb$ and light $M_H\simeq 200\gev$
or very heavy $M_H$.
The positive $\cba$ case allows the top-phobic lines
and thus $C=0$ line.
Similar to $\rr$ channel, the $\cosbma = 0.4$ case in Type I has two $C=0$ lines
which are narrowly split near the top-phobic line.
Near this top-phobic line $|\Delta C|$ is much enhanced
since $R$ is very suppressed.
Note that large signal rate region,
with small $M_H$ and some deviation from the top-phobic line,
has small $\dtc$.
The $\cosbma=0.1$ case in Type II allows one $C=0$ line,
below which $\dtc$ is negative.
For $M_H=200\gev$ and $\tb \simeq 5\sim10$,
the reduction is about 50\%.

 \begin{figure}[t]
  \centering
\vspace*{-0.2cm}
\includegraphics[width=0.96\textwidth]{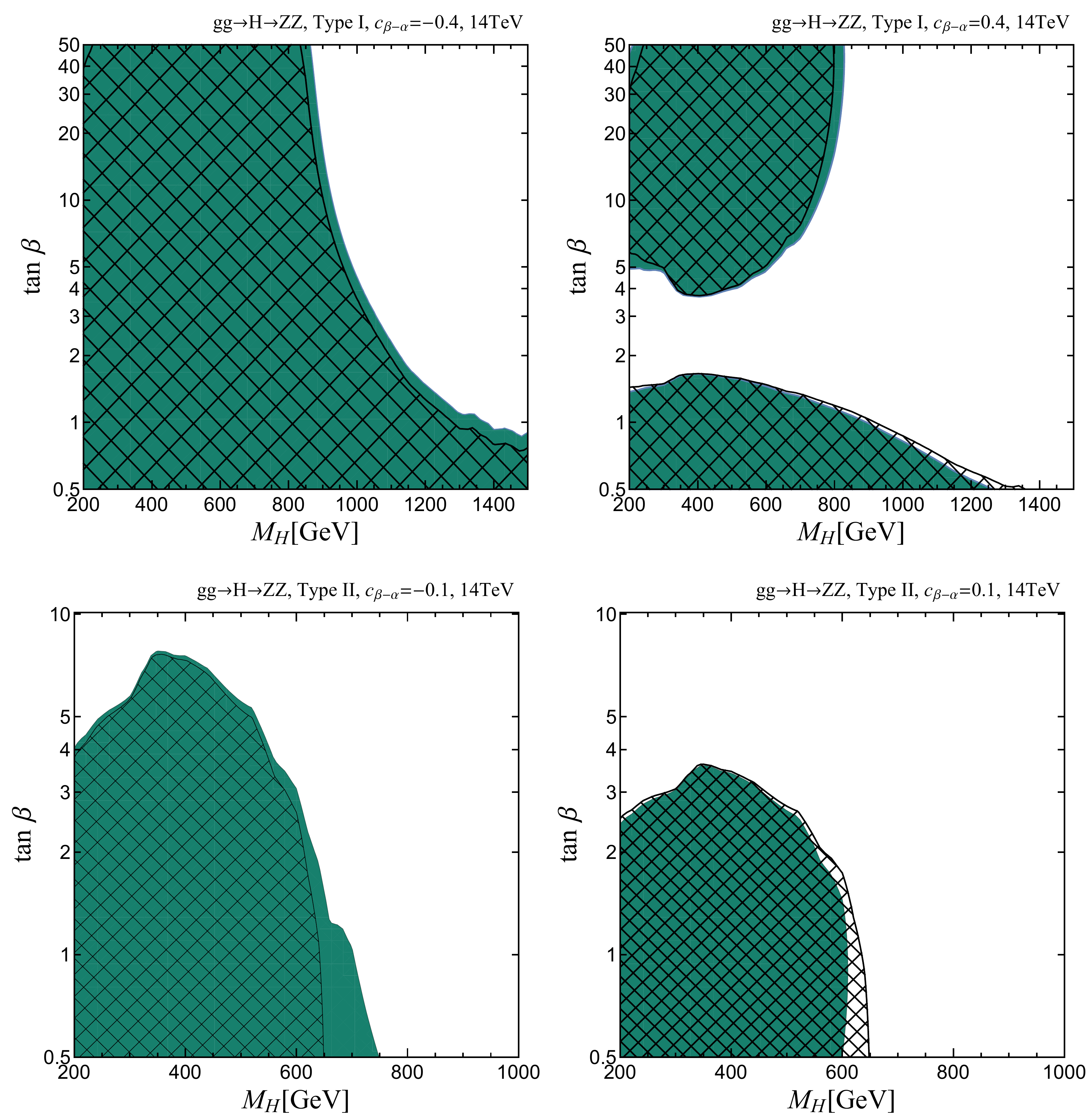}
  \caption{\baselineskip 3.5ex
  Expected exclusion regions from $ZZ$ resonance searches at the  14 TeV LHC with $300\invfb$, projected from the LHC 8 TeV results.
The green regions include interference effects
and  the hatched regions are without interferences. We used the mNWA to obtain these results assuming that a resonance shape is approximately a BW peak.
   }
  \label{fig:exc1}
\vspace{-0.2cm}
\end{figure}

We now investigate the interference effect
on the future experimental sensitivity for the heavy Higgs boson
 via $ZZ$ channel.
 We use the current experimental results
on the heavy neutral Higgs search in the $ZZ$ decay channel
with $20.3\invfb$ data at $8\TeV$ from ATLAS collaboration~\cite{Aad:2015kna}.
The non-observation of any significant excess above the background
is interpreted as an upper bound on $\sigma(gg\to H \to ZZ)$ as a function Higgs mass $M_H$.
We project the results for 14 TeV LHC with $300\invfb$ data
by assuming that the statistical error is dominant~\cite{Jung:2013zya,Djouadi:2015jea}.

 In Fig.~\ref{fig:exc1}, we show
 the  14 TeV LHC projections of the exclusion plots in $(M_H, \tb)$ plane
  for $300\invfb$ data.
  We set $\cosbma=\pm 0.4$ in Type I and $\cosbma=\pm 0.1$ in Type II.
  The hatched exclusion region is obtained by using naive NWA (without interference effect) while the green-colored exclusion region  by mNWA (with interference effect).
 For Type I, we extend the exclusion region up to $1.5\TeV$
 by applying a simple extrapolation.
 For $\cosbma =-0.4$
 the exclusion region covers the entire $\tb$ region up to $M_H=830\GeV$
 and  the interference effect enlarges the exclusion
 region by $20-30\GeV$,
 \ie constructive interference.
  For $\cosbma =0.4$, there are two separate exclusion
  regions divided by top-phobic line.
  The upper (lower) region shows constructive (destructive) interference effect.
Because of sizable top-quark Yukawa coupling above the top-phobic line, the large $\tb$ region  ($5<\tb<50$) up to $M_H=800\GeV$ also can be reached by LHC Run II data.
  As for type II, for $\cosbma = 0.1$ ($\cosbma =-0.1$)
  the interference effect is constructive (destructive).

 \begin{figure}[t]
  \centering
    \subfigure{\label{fig:DpDm:a}\includegraphics[width=0.46\textwidth]{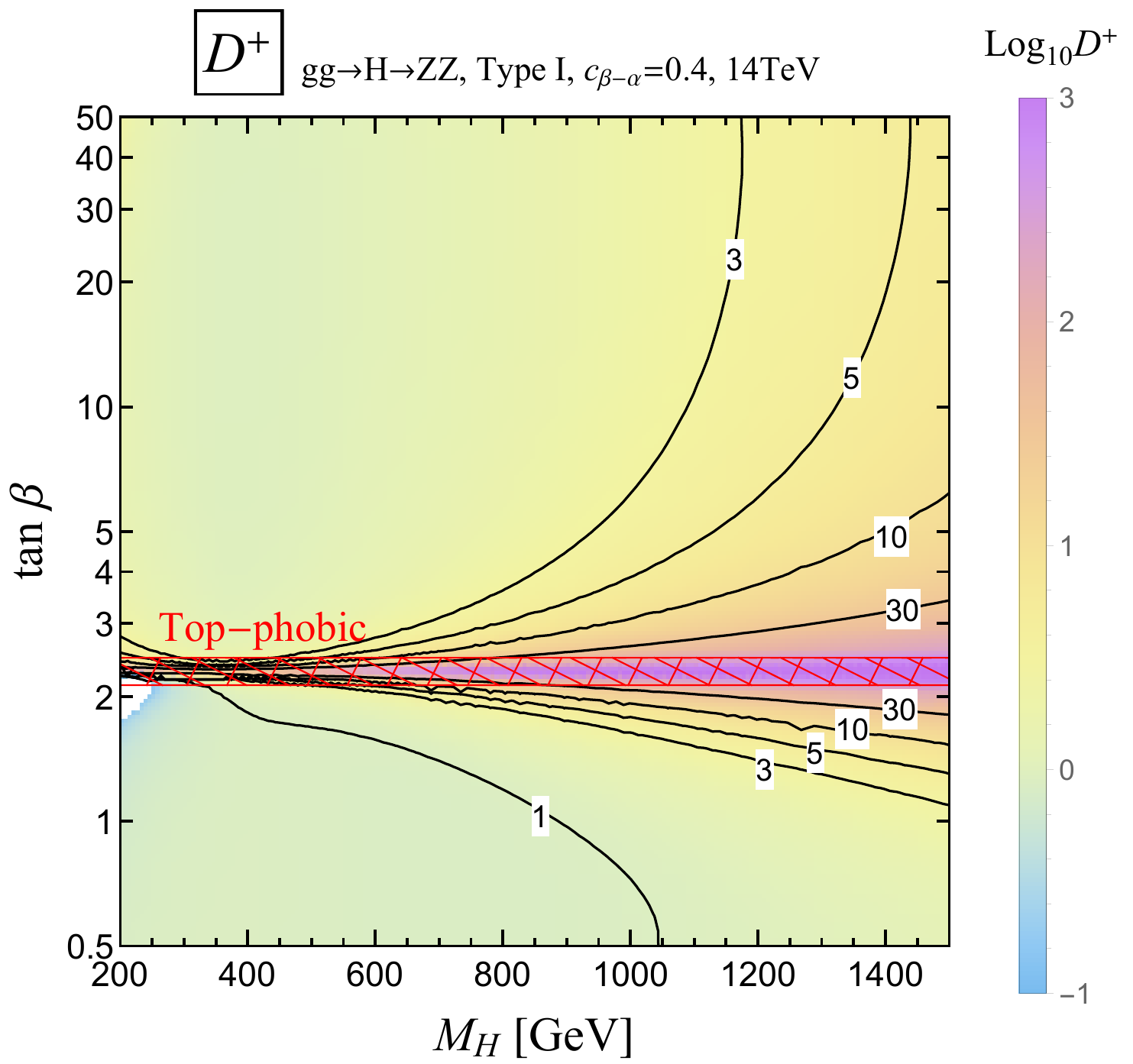}}
  \quad
  \subfigure{\label{fig:DpDm:b}\includegraphics[width=0.46\textwidth]{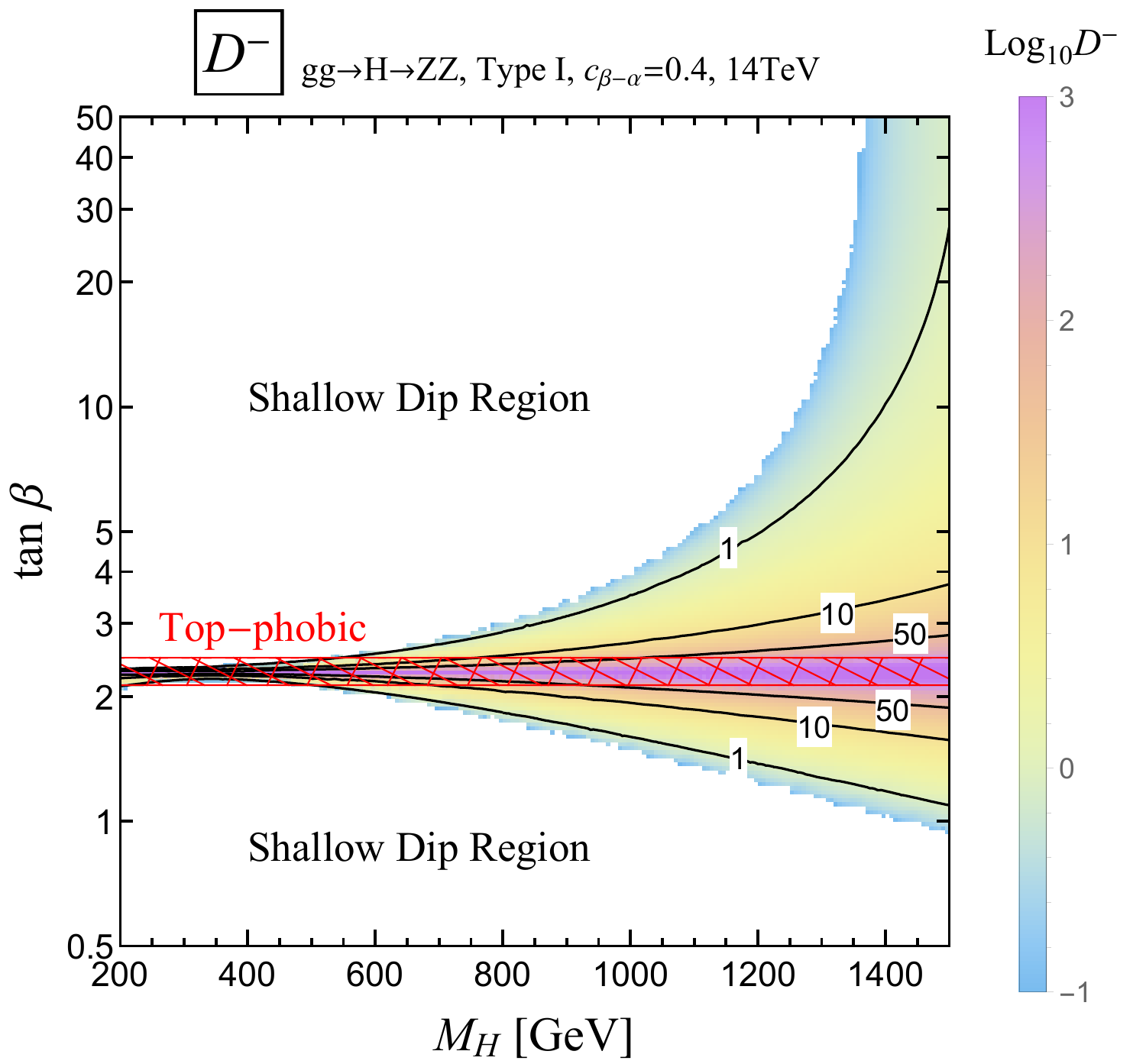}}\\
 \caption{\baselineskip 3.5ex
Contours of $D^+$ (left) and $D^-$ (right) in the $gg\to H \to ZZ$ process at the
$14\TeV$ LHC  for $\cosbma=0.4$ in Type I. $D^\pm$ are defined in \Eq{eq:dpm}. The red hatched region is around the top-phobic line such that $|{\hat y}^H_t|<0.03$.
}
  \label{fig:DpDm}
\end{figure}

Finally, we demonstrate the real-part interference effect
by showing the contours of $D^+$ and $D^-$ in $(M_H, \tb)$ plane:
see Fig.~\ref{fig:DpDm}.
We take the case for $\cosbma=0.4$ in Type I at $14\TeV$ LHC,
which brings about large real-part interference effects\footnote{
In other three cases,
the real-part interference are mostly minor. }.
The behaviors of $D^\pm$ are quite similar to $C$:
$D^\pm$ is large near top-phobic line where $R$ is small.
For $D^-$,
the uncolored region indicates that the dip is shallower than the half-maximum of
the BW peak.
Even in this shallow dip region,
$D^+$ can be up to 5,
which is attributed to highly asymmetric $m_{ZZ}$ distribution near $\hat s = M^2$.
Note that the corresponding  $|\dtc|$ factor is less than $\sim40\%$.
The real-part interference can be more important,
which is observable
if the detector resolution is good enough to
separate the peak from the dip.
Once the dip is deep enough (colored region),
both $D^{\pm}$ are larger than 1:
a clear dip-peak ($c_\phi > 0$) or peak-dip ($c_\phi < 0$) shape is expected.
The line-shape analysis can serve as another important probe for the heavy Higgs signal.

\section{Role of total decay width}
\label{sec:extra}

\begin{figure}[t]
  \centering
  \vspace*{-0.3cm}
  \includegraphics[width=0.96\textwidth]{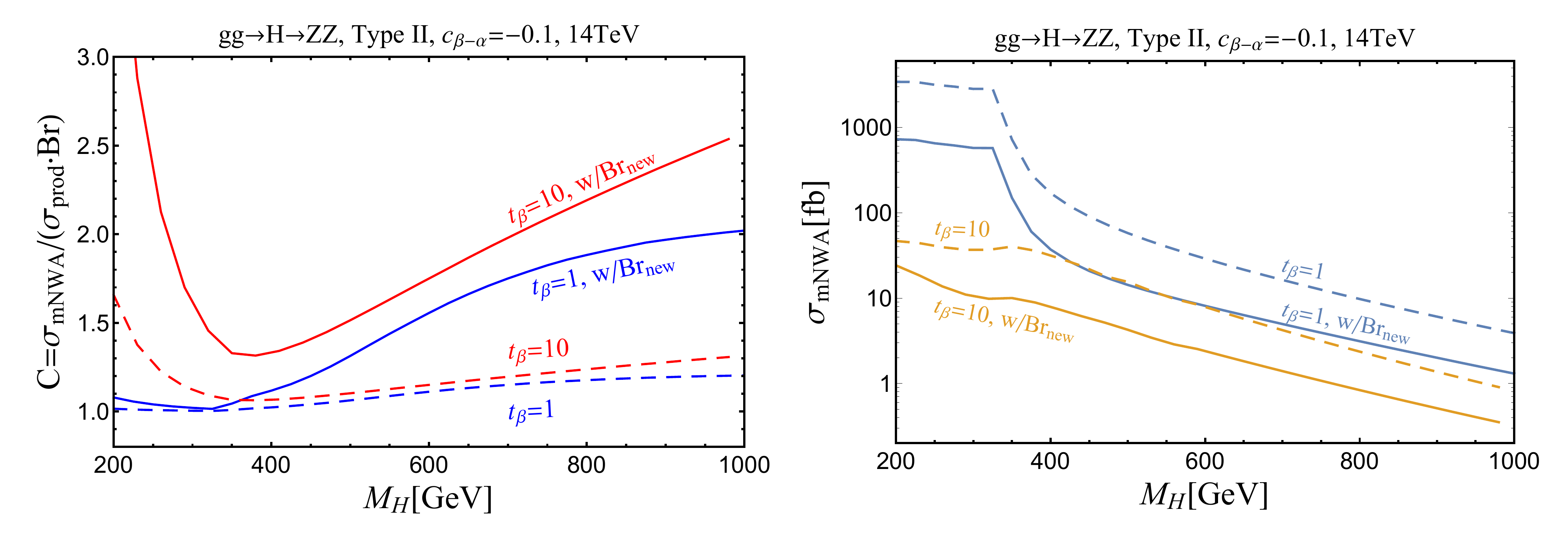}
  \caption{\baselineskip 3.5ex
  Plots of $C$ factor and $\sigma_{\rm mNWA}$ with respect to $M_H$ for $gg\to H\to ZZ$ with some parameter choices. Solid lines include new decay channel with $\br_{\rm new}=0.8$ while dashed lines do not.
 }
  \label{fig:exc2}
  \vspace{-0.3cm}
\end{figure}

So far, we have assumed that the heavy $H^0$ decays into the
SM fermions or gauge bosons just as the SM Higgs boson.
However, there exist additional decay channels like
$H\to hh, ZA,W^\pm H^\mp$ in the 2HDM.
As a benchmark scenario in Type II,
we consider
\begin{eqnarray}
&& \cosbma=-0.1,~\tb=1.8,~m_{12}=240\GeV,
\nn \\
&& M_H=550\GeV, ~M_A=350\GeV, ~M_{H^+}=350\GeV,
\end{eqnarray}
which satisfies the constrains from
$b\to s\gamma$~\cite{Mahmoudi:09,Hermann:2012fc},
$\Delta\rho$~\cite{Higgs:Hunters:Guide,Chankowski:1999ta}
as well as the stability and perturbativity~\cite{Kanemura:2004mg,2HDM}.
Then additional decay channels have sizable branching ratio:
$\br(H\to hh) = 0.004$,
$\br(H\to ZA) = 0.25$, and $\br(H\to W^\pm H^\mp) = 0.54$.
Or 2HDM can be extended to include a dark matter candidate $\chi$,
allowing a new decay channel of $H^0\to \chi\chi$.
The resulting increase in $w$ shall affect the interference effect:
$\dtc$ is enhanced while $\br(H\to ZZ)$ and thus
$\sigma_{\rm mNWA}$ are reduced.
Figure~\ref{fig:exc2} shows the plots of $C$ and $\sigma_{\rm mNWA}$
with $\br_{\rm new} = 0.8$ and some specific parameter choices in the $ZZ$ channel.
The magnitude of $|\Delta C|$ is significantly enhanced by a factor of about five.
As discussed before, $\sigma_{\rm mNWA}$ is much reduced.

\begin{figure}[t]
  \centering
    \vspace*{-0.2cm}
  \includegraphics[width=0.6\textwidth]{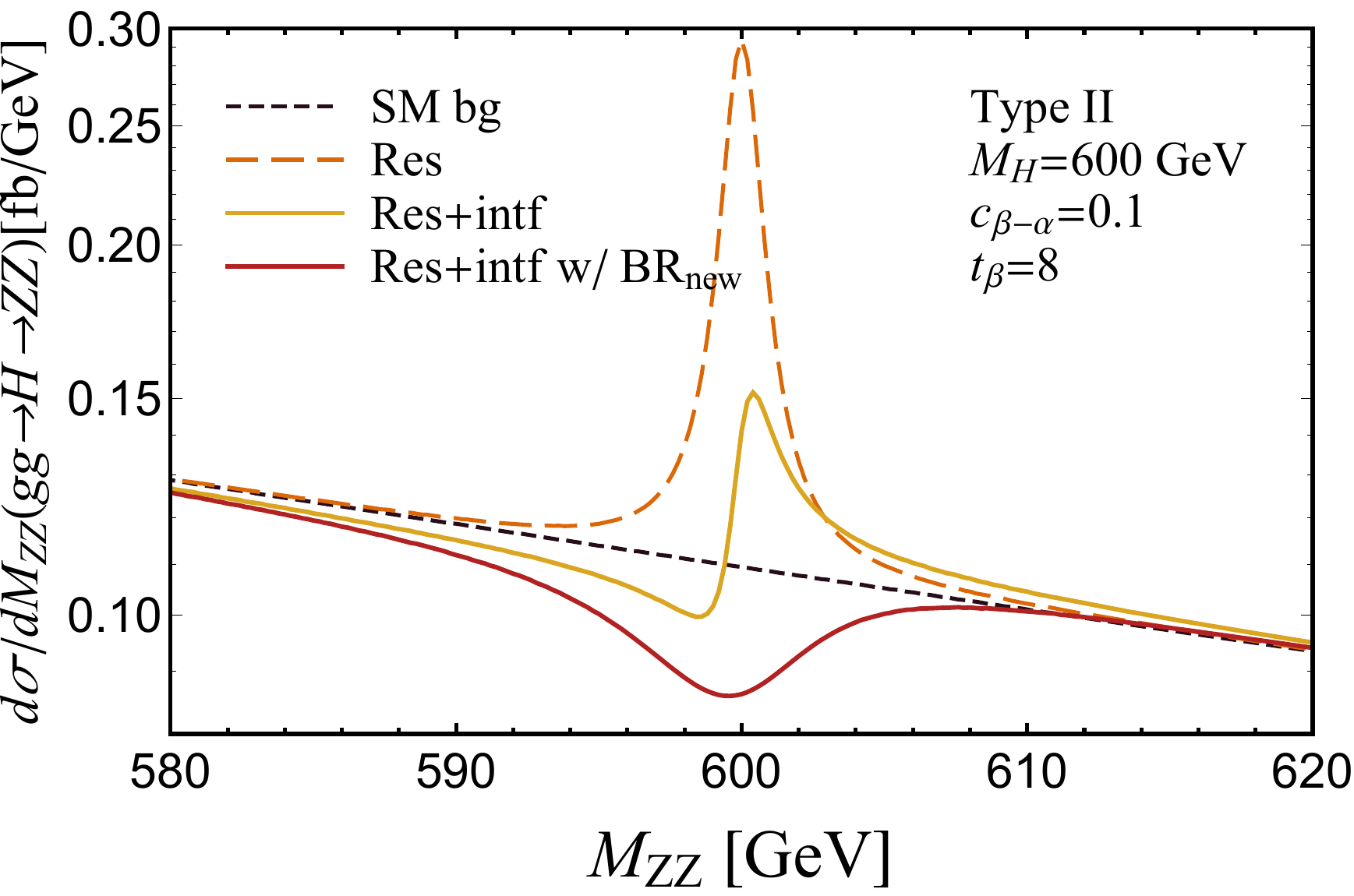}
  \caption{\baselineskip 3.5ex
Example $ZZ$ invariant mass distribution of $gg\to H\to ZZ$ for given parameter choice.
 We show several results for comparison: the full result with interferences accounted for (yellow-solid), the pure BW result without interferences (orange-dashed), the full result with additional decay modes $\br_{\rm new} = 0.8$ (red-solid), and the continuum background $gg \to ZZ$ alone (black-dashed).
    }
  \label{fig:dist}
    \vspace*{-0.2cm}
\end{figure}

 Finally, we illustrate how dramatically the resonance shape can be altered
 by changing the total width.
 Figure \ref{fig:dist} shows the $m_{ZZ}$ distribution
 for a benchmark point of $M_H=600\GeV$, $\cosbma=0.1$ and $\tb=8$
 in Type II,
 which yields $R=0.0063$, $\phi=-74^\circ$, and $\sigma \cdot \br \simeq 0.4\fb$.
If $H^0$ decays into the SM fermions and gauge bosons only,
the total decay width is small ($w=0.27\%$)
and $\dtc=-83\%$.
The resonance shape is  a suppressed dip-peak structure,
as denoted by the yellow solid line.
If we further allows a sizable branching ratio of a new decay channel
like $\br_{\rm new} = 0.8$,
the $m_{ZZ}$ distribution shape changes into a new form,
a pure dip (the red solid line).
Three $m_{ZZ}$ distribution lines clearly show that
measuring the invariant mass distribution would serve
as a multi-dimensional projection
to extract the information of $w$, $R$ and $\phi$.
This is to be compared with measuring the total signal rate only,
just a single dimensional projection.

\section{Conclusions}
\label{sec:conclusions}
We have studied the resonance-continuum interference effects of  the heavy neutral Higgs boson $H^0$ of the nearly-aligned Type I and II 2HDM in the $\rr$ and $ZZ$ channels. Even for the $H^0$ with a sub-percent width-to-mass ratio, the size and pattern of the interference effects can vary with underlying parameters and can be observably large.

For the parameter space where the LHC 14 TeV $ZZ$ resonance searches have sensitivities (particularly for small $t_\beta \lesssim 1$), the interference effects mildly modify the $ZZ$ signal rate by ${\cal O}(10)\%$ and change the exclusion reach on the $M_{H^0}$ by ${\cal O}(10)$ GeV. For example, the benchmark parameter $M_H=300\GeV$, $\cosbma=0.1$, and $\tb=10$ (yielding $\Gamma_H/M_H = 0.2\%$) modifies the $ZZ$ signal rate by $-30\%$ from the interference. In this parameter space, we approximately treated the resonance shape as a BW peak and used the mNWA to estimate a new exclusion reach. In other words, we could use the $C$-factor to approximately quantify the interference effects (hence, on the signal rate).

In other parameter space where the $ZZ$ signal rate is substantially smaller (such as near the top-phobic line), the LHC 14 TeV reach is absent but a resonance shape can be much more dramatically changed. Also for the $\rr$ channel, signal rates are smaller and the LHC 14 TeV reach is absent in all parameter space, but the signal rates can be typically modified by ${\cal O}(100)\%$ due to interference effects. For the same parameter choice as in the previous paragraph, the $\rr$ signal rate is modified by 400\%. In these cases, one shall carry out a more careful collider study of resonance searches including the interference effects. Rather than doing this in this paper, we suggested a few variables that can characterize a general resonance shape defined in Eqs.~(\ref{eq:hpm:wpm:general})-(\ref{eq:dpm}) and in Fig.~\ref{fig:D:definition}. It will be worthwhile carrying out a dedicated future collider study.

The interference effects also grow with the resonance width. To illustrate this, we also considered a case with extra non-standard decay modes of the heavy Higgs boson. The increased width and correspondingly reduced signal could induce more significant interference effects in both channels. For example, with 80\% new branching ratio, we find that the $ZZ$ signal rate can be enhanced by a factor 2-3, but a more careful collider study including the modification of a resonance shape will be needed in this case.

From varying sizes of the interference with different signal rates, we verify a general result that we have discussed in Ref.~\cite{Jung:2015gta}; the smaller the signal rate, the bigger the relative interference. The resonance-to-continuum ratio, approximately measured by $R$ in \Eq{eq:phi:definition}, is another factor; the smaller the ratio, the bigger the relative interference. This is why the $\rr$ channel experiences a bigger relative interference than the $ZZ$ channel does. Looking forward, a 100 TeV $pp$ collider and high-luminosity LHC data that can probe the parameter space with smaller signal rates will generically experience bigger interference effects.

\begin{acknowledgements}
The work of JS is supported by NRF-2013R1A1A2061331.
The work of YWY is supported by NRF-2012R1A2A1A01006053.
The work of SJ is supported by the US Department of Energy under contract DE-AC02-76SF00515.
We thank Convergence Computing team of National Institute for Mathematical
Sciences
for valuable comments in extracting data from experimental papers,
and KIAS Center for Advanced Computation for providing computing resources.
\end{acknowledgements}


\end{document}